%% file: appb.tex
\documentclass{appolb}
\usepackage{authblk}
\usepackage{float}

\usepackage[utf8]{inputenc}
\usepackage[T1]{fontenc}
\usepackage{graphicx}
\usepackage{grffile}
\usepackage{longtable}
\usepackage{wrapfig}
\usepackage{rotating}
\usepackage[normalem]{ulem}
\usepackage{amsmath}
\usepackage{textcomp}
\usepackage{amssymb}
\usepackage{capt-of}
\usepackage{hyperref}
\usepackage{tikz}
\usepackage{xcolor}
\input{Def-org.tex}

\usepackage{cite}
\usepackage[textwidth=169mm,textheight=229mm,hmarginratio=1:1,vmarginratio=1:1]{geometry}
\hypersetup{linktocpage,allcolors=black} 

\usepackage{graphicx}

\begin{document}
\title{Corrections to (pseudo)scalars decay into a fermion pair from gravitational torsion}
\author[1,2]{Felipe Rojas Abatte}
\author[3]{Jilberto Zamora-Saa\thanks{Corresponding Author:
  jzamorasaa@jinr.ru}}
\author[1,2]{Oscar Castillo-Felisola}
\author[1,2]{Bastian D\'iaz}
\author[1,2]{Alfonso R. Zerwekh}

\affil[1]{\CCTVal.}
\affil[2]{\UTFSM.}
\affil[3]{\DLNP.}

\maketitle

\begin{abstract}
  We study the contribution of the torsion-descendent four-fermion
  contact interaction to the decay width of a neutral (pseudo)scalar
  field into a fermion pair. This new interaction comes from the
  existence of gravitational torsion in models with extra
  dimensions. Additionally, we exemplify the formalism by studying two
  cases: first, the variation of the considered branching ratio of
  the Higgs in the context of the standard model, and second the
  proper variations of the scalar and pseudo scalar fields of the type
  II-1 two Higgs doublets model.
\end{abstract}
\PACS{04.50.Kd, 04.90.+e, 12.60.-i}

\section{Introduction}
\label{sec.intro}

The discovery of the Higgs boson in 2012 at the Large Hadron Collider
(LHC) \cite{Aaltonen:2012qt,Aad:2012tfa,Chatrchyan:2012ufa} not only
has completed the picture of the standard model (SM), but also has
opened the possibility of real existence of fundamental scalar fields
in nature. At the same time, some of the puzzles in the SM, such as
neutrino mass generation and dark matter, have stimulated the
scientific community to consider models with a larger scalar sector
\cite{jackiw73_dynam_model_spont_broken_gauge_symmet,cornwall73_spont_symmet_break_without_scalar_meson,Deshpande:1977rw,dimopoulos79_mass_without_scalar,eichten80_dynam_break_weak_inter_symmet,holdom81_raisin_sidew_scale,holdom85_techn,yamawaki86_scale_invar_hyper_model_dilat,appelquist86_chiral_hierar_flavor_chang_neutr_curren_hyper,appelquist87_chiral_hierar_from_slowl_runnin,filippi06_dark_matter_from_scalar_sector,LopezHonorez:2006gr,Grzadkowski:2009bt}.
Extended scalar sector are also \emph{predicted} by models like
supersymmetry
\cite{gunion86_higgs_boson_super_model,gunion86_higgs_boson_super_model_II,gunion88_higgs_boson_super_model},
some versions of strong electroweak symmetry breaking models
\cite{hapola12_pseud_golds_boson_phenom_minim_walkin_techn} and
non-minimal composite Higgs models
\cite{mrazek11_other_natur_two_higgs_doubl_model,bertuzzo13_compos_two_higgs_doubl_model,curtis16_theor_phenom_compos_doubl_model}.
Although no deviation from the SM has yet been observed, the LHCb
collaboration has reported anomalies in the Lepton Flavour
Universality violating ratios, \(R_K\) and \(R_{K^*}\). These
anomalies can be explained via models that include new heavy vector
and scalar bosons
\cite{capdevila17_patter_new_physic_trans_light_recen_data,ghosh17_explain_r_k_r_k_anomal,ko17_lhcb_anomal_b_physic_flavor}.

At the same time, there have been other extensions of the SM motivated
by the possible existence of more than three spatial dimensions
\cite{GSW1,GSW2,Polchinski1,Polchinski2,Tong-strings,Duff:1986hr,PopeKK}. In
these scenarios, it is tempting to consider (in the bulk) an extended
gravitational sector. Indeed, Einstein's theory of gravity, known as
General Relativity (GR), is now view as a low energy effective theory
of a (yet unknown) fundamental model, in particular due to the lack
of a consistent quantum version of the theory.\footnote{There are several attempts of quantize the gravitational
  interactions, see for example
  Refs. \cite{DeWitt:1967yk,DeWitt:1967ub,DeWitt:1967uc,Ashtekar:1986yd,Ashtekar:1987gu,Ashtekar:2004eh}. For
  a historical review, see Ref. \cite{Rovelli:2000aw}.} In an effort
to obtain a more fundamental theory of gravity, several
generalizations of GR have been proposed, from the \emph{minimal}
generalization of considering a metric compatible affine connection
\cite{Cartan1922,Cartan1923,Cartan1924,Cartan1925}, models which keep
the precepts of GR but in higher dimensions
\cite{lovelock1971einstein,Zumino:1985dp}, metric-affine theories
\cite{Hehl:1994ue}, affine theories
\cite{Eddington1923math,schrodinger1950space,Plebanski:1977zz,Krasnov:2011pp,Krasnov:2011up,poplawski13_affin_theor_gravit,Skirzewski:2014eta},
models with higher order in curvature and/or torsion
\cite{DeFelice:2010aj,Sotiriou:2008rp,Capozziello:2009mq,Fabbri:2012qr,Pagani:2015ema,Belyaev:1998ax,Belyaev:2007fn},
et cetera.

In this letter, we shall only consider the Cartan's generalization to
GR, usually called Einstein--Cartan theory of gravity (ECT), in which
the torsion turns out to be a non-dynamical field, and it can be
integrated out of the system. When the ECT of gravity is coupled with
fermionic matter, the integration of the torsion induces an effective
four-fermion contact interaction
\cite{Kibble:1961ba,Hehl:1976kj,Shapiro:2001rz,Hammond:2002rm}, whose
phenomenological effects can be observed at accelerators. It is
well-known that such induced effective interaction is strongly
suppressed because it is diminished by the of Newton's constant, or in
other words, by the inverse of the Planck mass squared. However, there
are scenarios with extra dimensions which achieve naturalness between
the electroweak, \(M_W\), and the (fundamental) gravitational scales,
\(M_*\), while the known Planck's mass, \(M_{\text{pl}}\), is an enhanced
effective gravitational scale \cite{ADD1,AADD,ADD2,RS1,RS2}. Therefore,
the suppression of the torsion descendent four-fermion interaction is
not as dramatic.

Among the phenomenological aspects which can be observed from the
induced four-fermion interaction one can name the following: several
cosmological problems
\cite{Poplawski:2010jv,Poplawski:2010kb,Poplawski:2011wj,Poplawski:2011xf,Fabbri:2012yg,Vignolo:2014wva},
the origin of fermion masses \cite{Castillo-Felisola:2013jva}, neutrino
oscillation
\cite{Capozziello:2013dja,cirilo-lombardo13_solar_neutr_helic_effec_new,Alvarez-Castillo:2016wve},
impose limits on extra dimensional model
\cite{Chang:2000yw,CCSZ,Castillo-Felisola:2014iia}, and changing
one-loop observable \cite{Lebedev:2002dp,Castillo-Felisola:2014xba}. 

A possible effect of this four-fermion interaction is to modify,
through one-loop effects, the decay width of generic (pseudo)scalar bosons
into a pair of SM fermions. The aforementioned is applicable, for
example, to the Higgs decay. This deviation from standard predictions
could be observed in principle, by means of precision measurements
performed in future lepton colliders as the International Linear
Collider (ILC) or the Compact Linear Collider (CLIC).

The letter is organized as follows. A brief review of the theoretical
setup is presented in Sec. \ref{sec.CEG}. In Sec. \ref{sec.1loop} we show
the one-loop corrections due to the effective interaction to the decay
width for a (pseudo)scalar boson decaying into a fermion pair. In
Sec. \ref{sec.Higgs} and Sec. \ref{sec.2HDM} we apply our result to the SM
Higgs, and to the (pseudo)scalars in the framework of two-Higgs doublets
model (2HDM), decaying into \(\tau^+\tau^-\) and \(b\bar{b}\) final
states. Finally, in Sec. \ref{sec.concl} we present a discussion of the
results and the concluding remarks.

\section{Effective interaction through gravitational torsion}
\label{sec.CEG}
The standard GR is interpreted as a field theory for the
metric. Since the field equations for the metric are of second order,
the approach is known as \emph{second order formalism}. However, even in
standard GR---where the connection is a metric potential---one can
treat the metric and the connection as independent fields, and their
field equations are then first order differential equations. This
latter approach is called \emph{first order formalism} or sometimes
\emph{Palatini's formalism} \cite{palatini1919deduzione}. Although the
Palatini's approach can be used with the metric and connection fields,
it is useful to consider an equivalent set of fields known as the
\emph{vielbein} (\(\vi{a}{\mu}\)) and the \emph{spin connection}
(\(\spi{\mu}{ab}{}\)),\footnote{The name ``spin connection'' is historical, and it is not
  necessarily related with the spin of the fields. For this reason some
  authors prefer to call it \emph{Lorentz connection}.} which encode the information of how to
\emph{translate} from the curved spacetime to the tangent space, and how
these tangent spaces are \emph{connected} with those of the neightbourhood
points.\footnote{The vielbein field ensures the validity of the equivalence
  principle.} The equivalence between the metric and the vielbein is given
by
\begin{equation}
  \label{eq.rel_metric_vielbein}
  g_{\mu\nu} = \eta_{ab} \vi{a}{\mu} \vi{b}{\nu}.
\end{equation}
Despite one can write down an explicit relation between the
Christoffel connection and the spin connection, we omit it. Instead,
we present the equations that define the torsion and curvature
two-forms,\footnote{We make extensive use of the formalism of differential forms \cite{Cartan-calc,Nakahara,Zanelli:2005sa,Zanelli:2016}.} i.e. the Cartan structure equations,
\begin{equation}
  \df[\vif{a}] + \spif{a}{b} \we \vif{b} = \Tf{a} 
  \quad
  \text{and}
  \quad
  \df[\spif{ab}{}] + \spif{a}{c} \we \spif{b}{c} = \Rif{ab}{}.
  \label{eq.Cartan_str}
\end{equation}
The vielbein and spin connection one-forms are defined as 
\begin{equation}
  \label{eq.vi_spi_forms}
  \vif{a} = \vi{a}{\mu}\de{x}^\mu
  \quad
  \text{and}
  \quad
  \spif{ab}{} = \spi{\mu}{ab}{}\de{x}^\mu,
\end{equation}
while the two-forms are written explicitly in components as
\begin{equation}
  \label{eq.tor_cur_forms}
  \begin{split}
    \tf{a} & = \frac{1}{2} \tors{\mu}{a}{\nu} \de{x}^\mu \we \de{x}^\nu
    = \frac{1}{2} \tors{m}{a}{n} \vif{m} \we \vif{n}
    \\
    \text{and}
    \\
    \rif{a}{b} & = \frac{1}{2} \ri{\mu\nu}{a}{b} \de{x}^\mu \we \de{x}^\nu
    = \frac{1}{2} \ri{mn}{a}{b} \vif{m} \we \vif{n}.
  \end{split}
\end{equation}
In the following, in order to distinguish among quantities in higher
or four dimensional spacetimes, we shall use the notation defined in
Refs. \cite{Castillo-Felisola:2013jva,Castillo-Felisola:2014iia,Castillo-Felisola:2014xba},
where hatted quantities refer to objects (or indices) lying in the
former, while unhatted quantities refer objects (or indices) lying in
the latter. Worth to mention, we denote by \(\ga^{*}\) the
four-dimensional chiral matrix and multi-index gamma matrices
represent the antisymmetrized product of gamma matrices, i.e.,
\(\gamma_{\mu_1\cdots\mu_n} = \gamma_{[\mu_1} \cdots \gamma_{\mu_n]}\).

As starting point, we consider the \(D\)-dimensional action which
includes ECT of gravity coupled minimally with Dirac
fields,\footnote{We assume that fermion masses are developed through
  the Higgs mechanism, so the is no need for considering nontrivial
  fundamental mass terms.}
\begin{equation}
  \begin{split}
    S & = \frac{1}{2\kappa^2}\int
    \frac{\epsilon_{\hat{a}_1\ldots\hat{a}_D}}{(D-2)!}\,\rif*{\hat{a}_1\hat{a}_2}{}
    \we \vif*{a_3}\we \ldots \we \vif*{a_D}
    \\
    & \quad - \frac{1}{2}\sum_{f}\int\left(\bar{\Psi}_f \hat{\gf} \we \st \hat{\cdf} \Psi_f -
    \hat{\cdf} \bar{\Psi}_f \we \st \hat{\gf} \Psi_f\right),
  \end{split}
  \label{eq.action}
\end{equation}
where \(\hat{\cdf}\) is the spinorial covariant derivative in a curved
spacetime, defined by\footnote{Hereon, multi-index gamma matrices represent the totally
  anti-symmetric product of gammas.}
\begin{equation}
  \begin{split}
    \hat{\cdf} \Psi & = \df[\Psi] + \frac{1}{4} \spif*{\hat{a}\hat{b}}{}
    \ga_{\hat{a} \hat{b}} \Psi,
    \\
    \hat{\cdf} \bar{\Psi} & = \df[\bar{\Psi}] - \frac{1}{4} \bar{\Psi} \spif*{\hat{a}\hat{b}}{}
    \ga_{\hat{a} \hat{b}} ,
  \end{split}
\end{equation}
the symbol \(\hat{\gf}\) denotes the contraction \(\ga_{\hat{a}}
\vif**{a}\), the \(\st\) stands for the Hodge \emph{star} map, and the subscript \(f\) stands for the fermion's flavour.

The field equation for the spin connection in Eq. \eqref{eq.action}
yields an algebraic equation for the components of the torsion,
\begin{equation}
  \frac{1}{2}\big[ \hat{\tor}_{\hat{b} \hat{c} \hat{a} }
    + \hat{\tor}_{\hat{b} \hat{a} \hat{c} }
    + \hat{\tor}_{\hat{a} \hat{b} \hat{c} } \big]
  \equiv \hcont{\hat{a} \hat{b} \hat{c} }{}{}
  = - \frac{\kappa^2}{4} \sum_f \bar{\Psi}_f \ga_{\hat{a} \hat{b} \hat{c}} \Psi_f,
  \label{eq.tor_eom}
\end{equation}
notice that the expression in the LHS is the contorsion, whose only
nontrivial component, from Eq. \eqref{eq.tor_eom}, is its totally
antisymmetric part.

The contorsion is the tensor which relates the \emph{affine} spin
connection with the torsion-less spin connection,
\(\spif-**{a}{b}\),
through the equation
\begin{equation}
  \spif**{a}{b} = \spif-**{a}{b} + \kf**{a}{b},
  \label{eq.rel_spi_cont}
\end{equation}
where the contorsion one-form is defined by \(\kf**{a}{b} =
\ko**{m}{a}{b} \vif**{m}\).

The advantage of Eq. \eqref{eq.tor_eom} been algebraic, is that it can
be substituted back into the action, allowing us to obtain an
effective, torsion-free action. The effective action includes GR
coupled minimally with the Dirac fields, plus an induced four-fermion
contact interaction, namely
\begin{equation}
  \Lag_{\text{4FI}} = \frac{\kappa^2}{32} \sum_{f_1,f_2} \big( \bar{\Psi}_{f_1} \ga_{\hat{a}
    \hat{b} \hat{c}}\Psi_{f_1} \big)  \big( \bar{\Psi}_{f_2} \ga^{\hat{a}
    \hat{b} \hat{c}}\Psi_{f_2}  \big).
  \label{eq.Lag4FI}
\end{equation}
In four dimensions---where \(\kappa^2 =
\frac{1}{M_{\text{pl}}^2}\)---the extra contact interaction is strongly
suppressed by the Planck mass, as anticipated. Therefore, this
effective interaction is negligible for any phenomenological effect.

Lately the phenomenological insight of scenarios with extra dimensions
has increased, boosted by works which solve the \emph{hierarchy
  problem},\footnote{One of the most outstanding proposals in the context of extra
  dimensions is the \(AdS/CFT\) correspondence (see for example
  Ref. \cite{Maldacena1,Maldacena2,natsuume15_ads_cft_dualit_user_guide}),
  which related different physical theories which live in different
  dimensions, reason why it is sometimes called \emph{holographic
    theory}. Nevertheless, we do not use the correspondence in this work.} i.e. the huge difference between the electroweak and
gravitational scales, through the introduction of a fundamental scale
of gravity, \(\kappa^{-1}_* = M_* \sim \si{\TeV}\), which gets enhanced in the
four-dimensional effective theory, up to the Planck scale
\cite{ADD1,AADD,ADD2,RS1,RS2}.

Within the framework of model with extra dimensions, the coupling
accompanying the effective four-fermion interaction in
Eq. \eqref{eq.Lag4FI}, should be replaced from \(\kappa\) to
\(\kappa_*\), which permits---in principle---to obtain some particle
physics phenomenology from the gravitational induced term.

We restrict ourselves to consider a single extra dimension in the rest
of the paper. As a first step, we decompose the induced four-fermion
interaction in terms of four-dimensional quantities, using that the
five-dimensional Clifford algebra admits the same representation as
the one in four dimensions. Therefore,
\begin{equation}
  \big(\ga_{\hat{a} \hat{b} \hat{c}}\big) \big(\ga^{\hat{a} \hat{b}
    \hat{c}}\big)
  = \big(\ga_{a b c}\big)\big(\ga^{a b c}\big) + 3 \big(\ga_{a b *}\big)\big(\ga^{a b *}\big).
\end{equation}
Hence, the interaction in Eq. \eqref{eq.Lag4FI} rises an axial--axial
and a tensor-axial--tensor-axial interactions
\cite{Castillo-Felisola:2013jva}
\begin{equation}
  \begin{split}
    \Lag_{\text{eff}}
    & = \frac{3 \kappa_*^2}{16} \sum_{f_1,f_2} 
    \big( \bar{\Psi}_{f_1} \ga_{a}\ga^* \Psi_{f_1} \big)
    \big( \bar{\Psi}_{f_2} \ga^{a}\ga^* \Psi_{f_2} \big)
    \\
    & \quad + \frac{3 \kappa_*^2}{32} \sum_{f_1,f_2}
    \big( \bar{\Psi}_{f_1} \ga_{a b}\ga^* \Psi_{f_1} \big)
    \big( \bar{\Psi}_{f_2} \ga^{a b}\ga^* \Psi_{f_2} \big)
  \end{split}
  \label{eq.Eff4fi}
\end{equation}
where \(\ga^*\) is the chiral matrix in four dimensions.

\section{One-loop correction of decay width for a (pseudo)scalar into a pair of fermions}
\label{sec.1loop}
\begin{figure}[ht]
  \centering
  \begin{tikzpicture}[line width=1.0 pt, scale=1.3]
    \begin{scope}[shift={(0,0)}]
      \draw[scalar] (-180:1)--(0,0);
      \node at (-180:1.2) {\Large $\varphi$};
      \draw[fermionbar] (0,0) arc (180:05:.7);
      \draw[fermion] (0,0) arc (180:355:.7);
      \draw[fermionbar] (1.4,0.05) -- (2.5,.8);
      \draw[fermion] (1.4,-0.05) -- (2.5,-.8);
      \node at (2.7,0.8) {\Large $\overline{f}$};
      \node at (2.7,-0.8) {\Large $f$};
      \node at (0.7,-1.5) {(a)};
    \end{scope} 
    \begin{scope}[shift={(5,0)}]
      \draw[scalar] (-180:1)--(0,0);
      \node at (-180:1.2) {\Large $\varphi$};
      \draw[fermionbar] (0,0) arc (180:0:.7);
      \draw[fermion] (0,0) arc (180:360:.7);
      \draw[fermionbar] (1.5,0) -- (2.5,.8);
      \draw[fermion] (1.5,0) -- (2.5,-.8);
      \node at (2.7,0.8) {\Large $\overline{f}$};
      \node at (2.7,-0.8) {\Large $f$};
      \node at (0.7,-1.5) {(b)};
    \end{scope} 
  \end{tikzpicture}
  \caption{Scalar to fermion pair through the four-fermion interaction in \mbox{s-channel} (a) and \mbox{t-channel} (b).}
  \label{fig.s-t}
\end{figure}
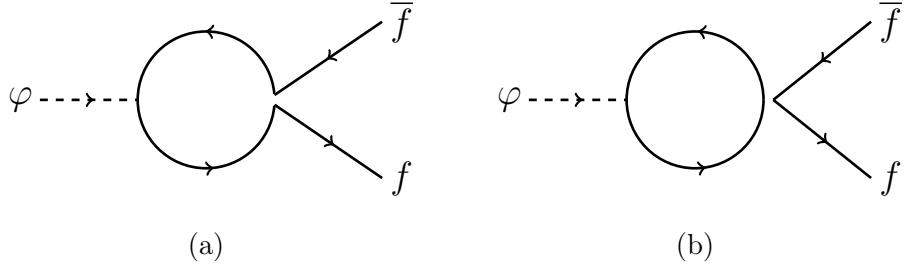

The splitting of the effective interaction, Eq. \eqref{eq.Eff4fi}, can
be written in terms of current--current interactions, as shown in
Ref. \cite{GonzalezGarcia:1998ay},
\begin{equation}
  \Lag_{\text{eff}} =
  \frac{3 \kappa_*^2}{16} \sum_{f_1, f_2} \big( J_{a\, f_1}^* \big)
  \big( J^{a*}_{f_2} \big)
  + \frac{3 \kappa_*^2}{32} \sum_{f_1, f_2} \big( J_{ab\, f_1}^* \big)
  \big( J^{ab*}_{f_2} \big).
  \label{eq.new4fi}
\end{equation}
There are two different contributions to the \(\varphi \to f \bar{f}\)
process, which will be called \emph{s-channel} (see Fig. \ref{fig.s-t}
(a)) and \emph{t-channel} (see Fig. \ref{fig.s-t} (b)) respectively. It
is worth to mention that in order to obtain chiral fermions in the
effective four-dimensional theory, an orbifold condition must be
imposed in the extra dimension \cite{Lebedev:2002dp}, and such condition
avoid the presence of tensor-axial--tensor-axial currents in
Eq. \eqref{eq.new4fi}. Therefore, in the below analysis only the induced
axial--axial currents will be considered.

We assume that the (pseudo)scalar fields couple to fermions through
generic Yukawa interactions, whose couplings are not necessarily
proportional to the final state fermion mass. Further, we assume that
the scalar field \(\varphi_s\) is CP-even, and the pseudo-scalar field
\(\varphi_p\) is CP-odd. Then, our Lagrangian contains the terms
\begin{equation}
  \Lag = \sum_f y_s^f \varphi_s \bar{\psi}_f\psi_f
  + \imath \sum_f y_p^f \varphi_p (\bar{\psi}_f \gamma^* \psi_f)
   \label{eq.Lag} 
\end{equation}
where \(y_{s,p}^f\) are real and arbitrary constants, and the \(f\)
index runs for each SM fermion, without considering
neutrinos. On the other hand, a (pseudo)scalar field decays into a
fermion pair through a current of the form
\begin{equation}
  J = \bar{u}_f(\vec{p}) \big( S + \imath P \ga^* \big) v_f(\vec{p}')
  \label{eq.Scur}
\end{equation}
where \(S\) and \(P\) are the scalar and pseudo-scalar form
factors. According to the current in Eq. \eqref{eq.Scur}, the decay
width of a (pseudo)scalar particle into a fermion pair at tree level
is given by
\begin{equation}
  \Gamma(\varphi \to f \bar{f} )
  =
  N_c\frac{M_{\varphi}}{8\pi} \sqrt{ 1 - \frac{4 m_f^2}{M_\varphi^2} }
  \left( (y_s^f)^2 S^2 \left(1 - \frac{4 m_f^2}{M_\varphi^2} \right) +
  (y_p^f)^2 P^2 \right),
  \label{eq.scalar_width}
\end{equation}
where \(M_\varphi\) is the mass of the (pseudo)scalar, \(m_f\) is the
fermion mass in the final state of the process and \(N_c\) is the colour
factor, which in the case of decay into quarks it will take the value
\(N_c = 3\).\footnote{We have cross-checked our calculations using the Mathematica package
``FeynCalc'' \cite{Kublbeck:1992mt}.}

It is worth noticing that the structure of the induced four-fermion
interaction in Eq. \eqref{eq.new4fi}, the t-channel Feynman
diagram---see Fig. \ref{fig.s-t} (b)---does not contribute to the decay width of the scalar nor pseudoscalar field. In the former,
the trace of the product of Dirac matrices
\begin{equation*}
  \Tr\left[(\not{p}_1 +\not{p}_2 - \not{q} - m_i)(\not{q} - m_i)\gamma^\mu\gamma^*\right]
\end{equation*}
vanishes identically due to the presence of the chiral matrix ($\ga^*$), while in the latter,
although the trace of the amplitude is nontrivial, the Feynman
integral results to be zero.

Next, we want to estimate the order of the correction to the decay
width induced by the four-fermion interaction described above. For
that end, we assume that the fundamental scale of gravity \(M_*\) is of
the order of the \emph{new physics} scale \(\Lambda\) ($M_* \sim \Lambda $). Therefore, although
our result comes from generic models with an extra dimension, we hide
the details of the model, such as the size of the extra dimension and
the embedding of the four-dimensional spinors into the
five-dimensional ones, within this new scale of physics.

The one-loop corrections to the current in Eq. \eqref{eq.Scur}, $\delta J$, through
the scalar field  decay into two fermions, considering the
effective four-fermion interaction is
\begin{equation}
  \delta J = \bar{u}_f(\vec{p}) \big( \delta S ) v_f(\vec{p}') \quad
  \text{with} \quad \delta S = - \frac{3}{32} \frac{1}{\Lambda^2} \big( M_\varphi^2 - 2 m_f^2 \big) \log\Big( \frac{\Lambda^2}{M_\varphi^2} \Big),
\end{equation}
while for the pseudoscalar is
\begin{equation}
  \delta J = \bar{u}_f(\vec{p}) \big( \imath P \ga^* \big) v_f(\vec{p}') \quad
  \text{with} \quad \delta P = - \frac{3}{32} \frac{1}{\Lambda^2}
  \left( M_\varphi^2 - 6 m_f^2 \right)
  \log\left( \frac{\Lambda^2}{M_\varphi^2} \right).
\end{equation}

Keeping the original coupling (tree level) and accounting for
CP-invariance. These results generate corrections to the variation of
the decay width of the form
\begin{equation}
  \delta \Gamma_{\text{4FI}}^{\text{S}}
  = -\frac{3}{128}
  \frac{N_c(y_s^f)^2M_\varphi}{\pi\Lambda^2}
  \left( M_\varphi^2 - 2 m_f^2 \right)
  \left( 1 - \frac{4m_f^2}{M_\varphi^2} \right)^{3/2}
  \log\left( \frac{\Lambda^2}{M_\varphi^2} \right) ,
  \label{eq.gen_Gamma_S}
\end{equation}
and
\begin{equation}
  \delta \Gamma_{\text{4FI}}^{\text{P}} =
  -\frac{3}{128} \frac{N_c(y_p^f)^2 M_\varphi}{\pi\Lambda^2}
  \left(M_\varphi^2 - 6 m_f^2\right)
  \left( 1 - \frac{4m_f^2}{M_\varphi^2} \right)^{1/2}
  \log\left( \frac{\Lambda^2}{M_\varphi^2} \right) .
  \label{eq.gen_Gamma_P}
\end{equation}

In these two cases the original result is a function of the
Passarino-Veltman integrals, however we have written the expressions
with the explicit logarithmic dependence on the scale \(\Lambda\).

\section{Standard Model Example: correction to Higgs decay into a pair of fermions}
\label{sec.Higgs}

Now we focus on special case of the Higgs boson
decay. As mentioned above, only the s-channel diagrams contribute to
the variation of the Higgs decay width. Furthermore, due to the fact SM Higgs is a scalar particle the quantities \(S\) and \(P\) in Eq. \eqref{eq.scalar_width} are
one and zero, respectively. Since the torsion induced four-fermion
interaction comes from the kinetic term, although the dimensional
reduction induces a Kaluza--Klein tower in the effective particle
spectrum, indisputably the fermion around the loop has the same
flavour as the outgoing particles. Therefore, none of the particles in
the Kaluza--Klein tower enter in the analysis. Then, the correction to the
variation of the Higgs decay width is
\begin{equation}
  \delta \Gamma_{\text{4FI}}(h \to f \bar{f})
  = -\frac{3}{512} \frac{g^2 m_f^2
    M_h}{\pi M_W^2\Lambda^2}
  (M_h^2 - 2m_f^2)
  \left(1 - \frac{4 m_f^2}{M_h^2}\right)^{3/2}
  \log\left(\frac{\Lambda^2}{M_h^2}\right).
  \label{higgs-width}
\end{equation}

\begin{figure}[H]
  \includegraphics[width=.49\textwidth]{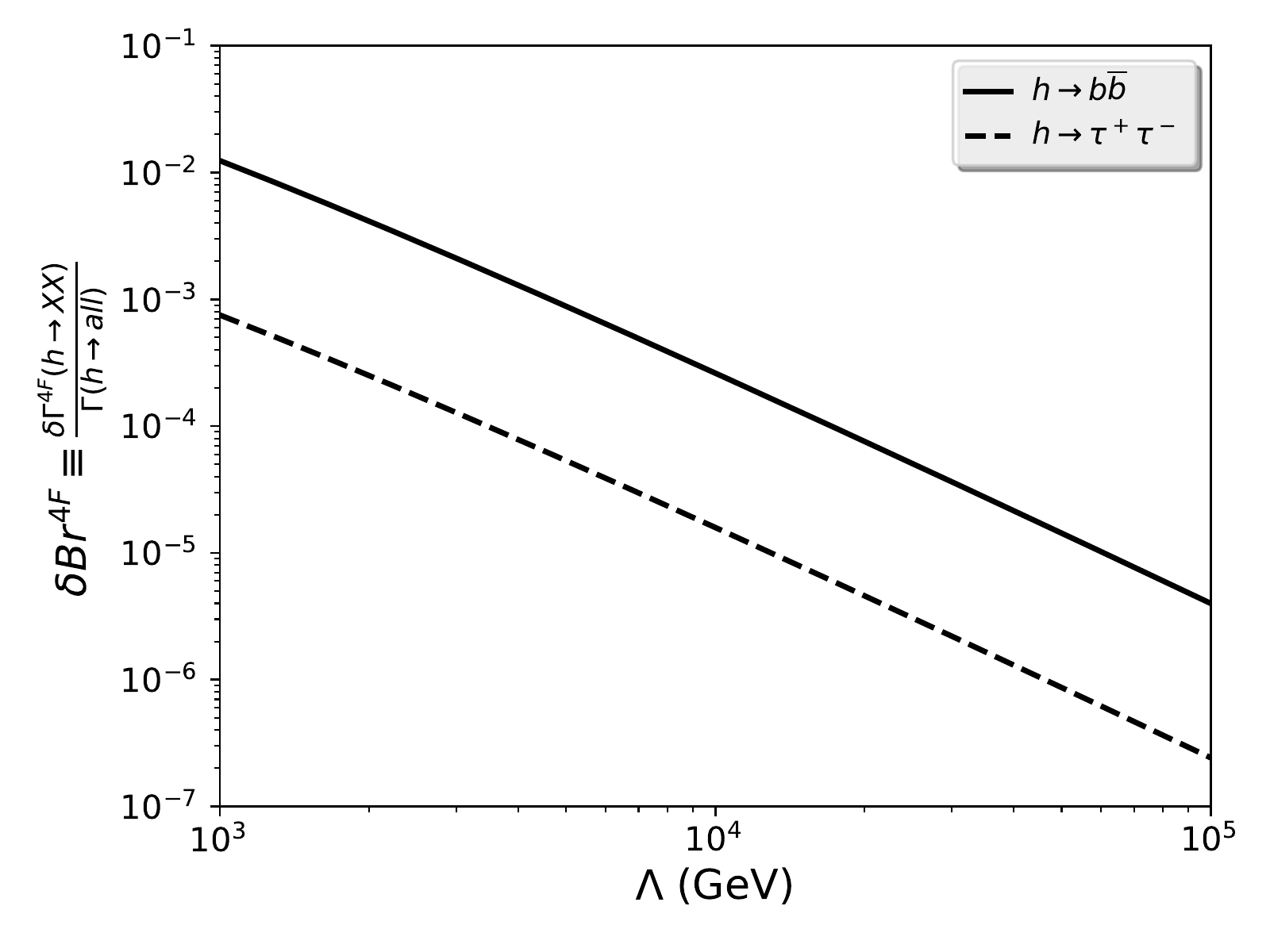}
  \includegraphics[width=.49\textwidth]{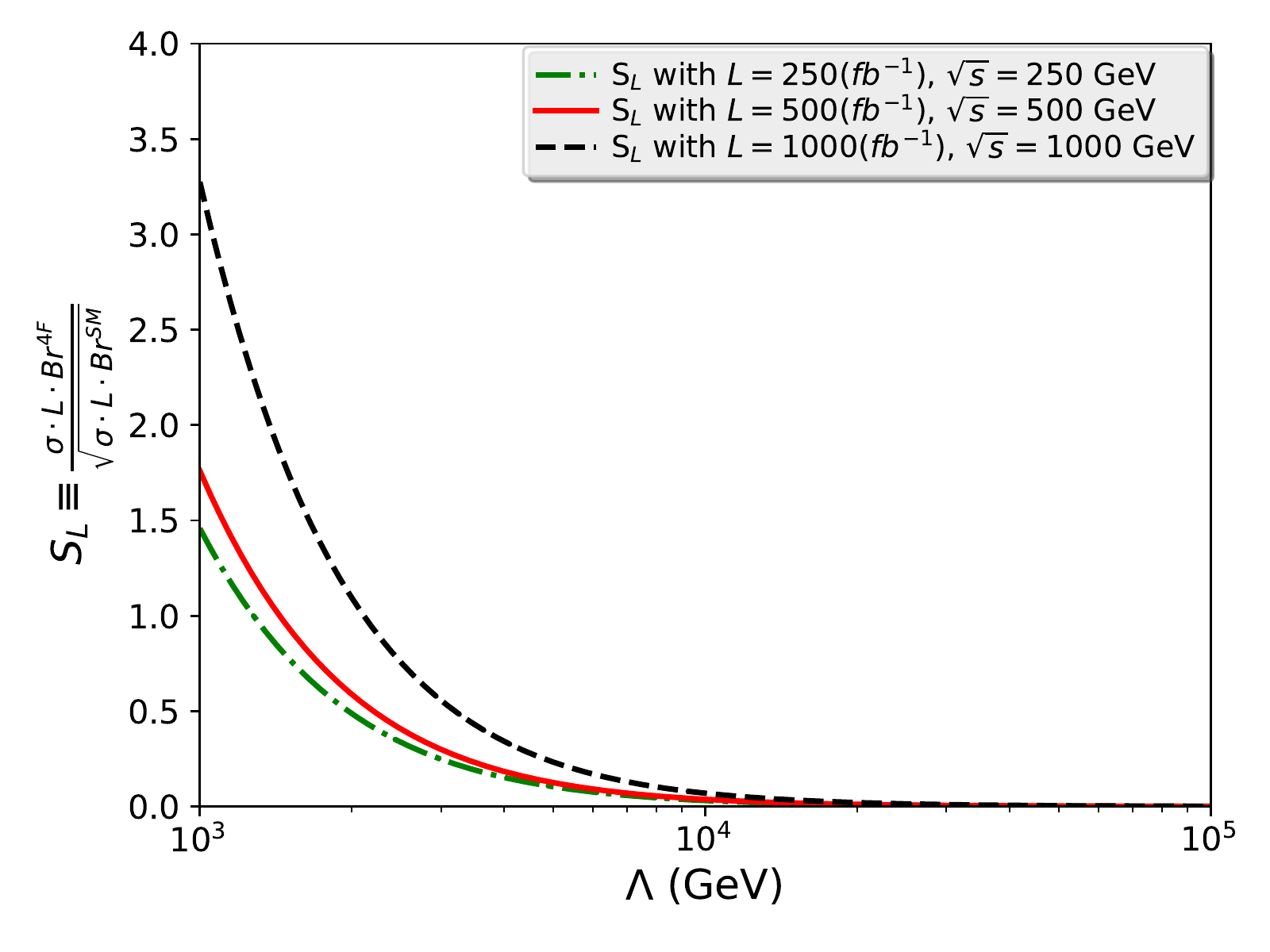} 
   \newline\newline
  \hspace*{\stretch{1}} (a)\hspace{\stretch{2}}(b)\hspace{\stretch{1}}
  \caption{(a) Variation of the Higgs boson branching ratio \( \delta
    Br^{4F} \) due to the 4-fermion interaction as a function of the
    new physics scale $\Lambda$. The dashed line denotes $h \rightarrow \tau^+ \tau^-$
    decay and the solid one $h \rightarrow  b \overline{b}$ decay channel. (b) Expected
    Significance level ($S_L$) at ILC.}
  \label{fig.Hdt}
\end{figure}

We will focus on Higgs decays into both \(\tau^+ \tau^-\) and \(b
\bar{b}\), which are the main fermionic decay modes, in order to
estimate the size of the effects and compare these corrections
with the total decay width predicted by the SM. In
Fig. \ref{fig.Hdt} (a) we show the correction on the Higgs branching ratio into fermion pairs as function of the gravitational scale.
For fundamental gravitational scales as low as \SI{1}{\TeV}, the
correction induced by the torsion interaction is about
\SI{1.024}{\percent} for the decay channel \(h \to b\bar{b}\), while for
the process \(h \to \tau^+ \tau^-\) it decreases to
\SI{0.075}{\percent}. As it is expected, for higher gravitational scales
the correction decreases due to the quadratic suppression ($\Lambda^{-2}$) in Eq.~\eqref{higgs-width}. 

Although the dominant Higgs branching fractions come from
\(h \to b \bar{b}\) and \(h \to \tau^+ \tau^-\), the LHC coupling
precision capabilities are not good enough to resolve what we are
interested in, due to the presence of QCD backgrounds~\cite{PDG,peskin12_compar_lhc_ilc_capab_higgs}.
However, the \(b\bar{b}\) signal channels may be more visible at future Higgs factories, such as the
ILC~\cite{weiglein06_physic_inter_lhc_ilc,peskin12_compar_lhc_ilc_capab_higgs,bechtle14_probin_stand_model_with_higgs} or CLIC~\cite{Battaglia:2004mw,battaglia03_study_higgs_sector_at_clic,asner03_higgs_physic_with_collid_based_clic,abramowicz16_higgs_physic_at_clic_elect}, where the QCD background is reduced. Then, it is
expected higher precision measurements in the Higgs sector, allowing 
to explore deeply the couplings and decay width, and therefore being 
able to
measure deviations in the Higgs decay width, eventually as low as our
results.

Keeeping in mind the aforementioned conditions, we estimate 
the expected significance
level (\(S_L\)) at the ILC, coming from the four-fermion
interaction. Such estimation reads
\begin{equation}
  S_L = \frac{\sigma \times L \times Br^{4F}(\Lambda)}{\sqrt{\sigma \times L \times Br^{SM}}} \epsilon_{f},
  \label{eq.Sign}
\end{equation}
where \(\sigma\) is the production cross section of the Higgs boson
via Higgsstrahlung \(\sigma = \sigma(e^+e^- \rightarrow h Z)\) and
vector boson fusion \(\sigma(e^+e^- \to \nu \bar{\nu} h)\), \(L\) is
the expected luminosity for each run, and \(\epsilon_f\) is the 
signal selection efficiency for the \(f\) channel, which is 
approximatelly \(\epsilon_f \simeq 0.3\) in both of the considered 
channels~\cite{Asner:2013psa,kawada15_study_measur_precis_higgs_boson}.
 As shown in Fig. \ref{fig.Hdt} (b), as the
gravitational scale increases, the expected significance in the number
of events---due to the torsion---decreases. It tell us that the effect
is observable at ILC only if \(\Lambda \sim \SI{1}{\TeV}\). Therefore,
if ILC does not see a significant excess of events in both \(b\bar{b}\)
and \(\tau^+\tau^-\) channels this energies scales, either the scale of
gravity is much bigger than these energy scales, or ETC gravity is not
coupled minimally to fermions.

However, recent analysis on the constraints imposed by the torsion
induced four-fermion interaction on the \(Z\) boson decay (see
Refs. \cite{Lebedev:2002dp,Castillo-Felisola:2014xba}), the strongest
limit is \(\Lambda \simeq \SI{30}{\TeV}\). Given this stringent limit,
the correction to the decay width of the Higgs drops to approximately
\SI{3.3e-3}{\percent} and \SI{2.2e-4}{\percent} for bottom and tau
pairs respectively. Such limits are unlike to be measure in current
experiments, but could be reached at future Higgs factories.

\section{Beyond Standard Model Example: 2HDM}
\label{sec.2HDM}
The 2HDM has in its physical spectrum two neutral scalar ($h^0,H^0$), one pseudo-scalar ($A^0$), and two charged bosons \((H^\pm)\), see
for example Ref. \cite{chang13_compr_study_two_higgs_doubl}. We focus on
the coupling between neutral bosons and SM fermions. The
parametrization of the Yukawa interactions in this context is
\begin{equation}
  \mathcal{L}_{\text{Yuk}} = -\sum_f \frac{m_f}{v} \big(\hat{y}_f^h \bar{f}fh^0 +
  \hat{y}_f^H \bar{f}fH^0 - \imath \hat{y}_f^A \bar{f}\gamma^5 f A^0 \big)
  \label{Yuk2HDM}
\end{equation}
where the constants \(\hat{y}_f^{h,H,A}\) are real numbers which depend on the
specific model, and \(v\) is the vaccum expectation value of the Higgs
field. There exist a diversity of forms of the 2HDM (Type I, II, X and
Y), but we shall consider the type II in its first scenario, called Type
II-1, which has the best fits to the observed data. In this scenario,
the \(h^0\) state match with the observed \SI{126}{\GeV} resonance
observed \(h\) at LHC, then \(h^0 = h\), and the \(\hat{y}_f^h\) measure the
deviation at tree level in the coupling between the Higgs and the SM
fermions. The other neutral scalars are heavy than the corresponding
Higgs boson and the coupling constants \(\hat{y}_f^{H,A}\) are determined by the Type II-1 model \cite{chang13_compr_study_two_higgs_doubl}.

\subsection{Corrections to the Higgs decay width in Type II-1 2HDM}
\label{sec:org3bd761d}

We compare the corrections to the Higgs decay width, induced by the
torsion-descendent four-fermion interaction, in two possibles
submodels: the constrained by
flauvor-physics and the unconstrained
\cite{chang13_compr_study_two_higgs_doubl}. 
We summarise the
values of the Yukawa couplings in both submodels in the Table~\ref{tab.yuk}.

\begin{table}[htbp]
  \caption{\label{tab.yuk}
    Yukawa couplings to up-type quark, to down-type quark, and to charged leptons for both submodels.}
  \centering
  \begin{tabular}{c|c|c}
    Yukawa coupling & Constrained & Unconstrained\\
    \hline
    \(\hat{y}^h_u\) & 1.28 & 1.05\\
    \(\hat{y}^h_d\) & -0.91 & -0.99\\
    \(\hat{y}^h_l\) & -0.91 & -0.99
  \end{tabular}
\end{table}

\begin{figure}[!htb]
  \begin{center}
    \includegraphics[width=.6\textwidth]{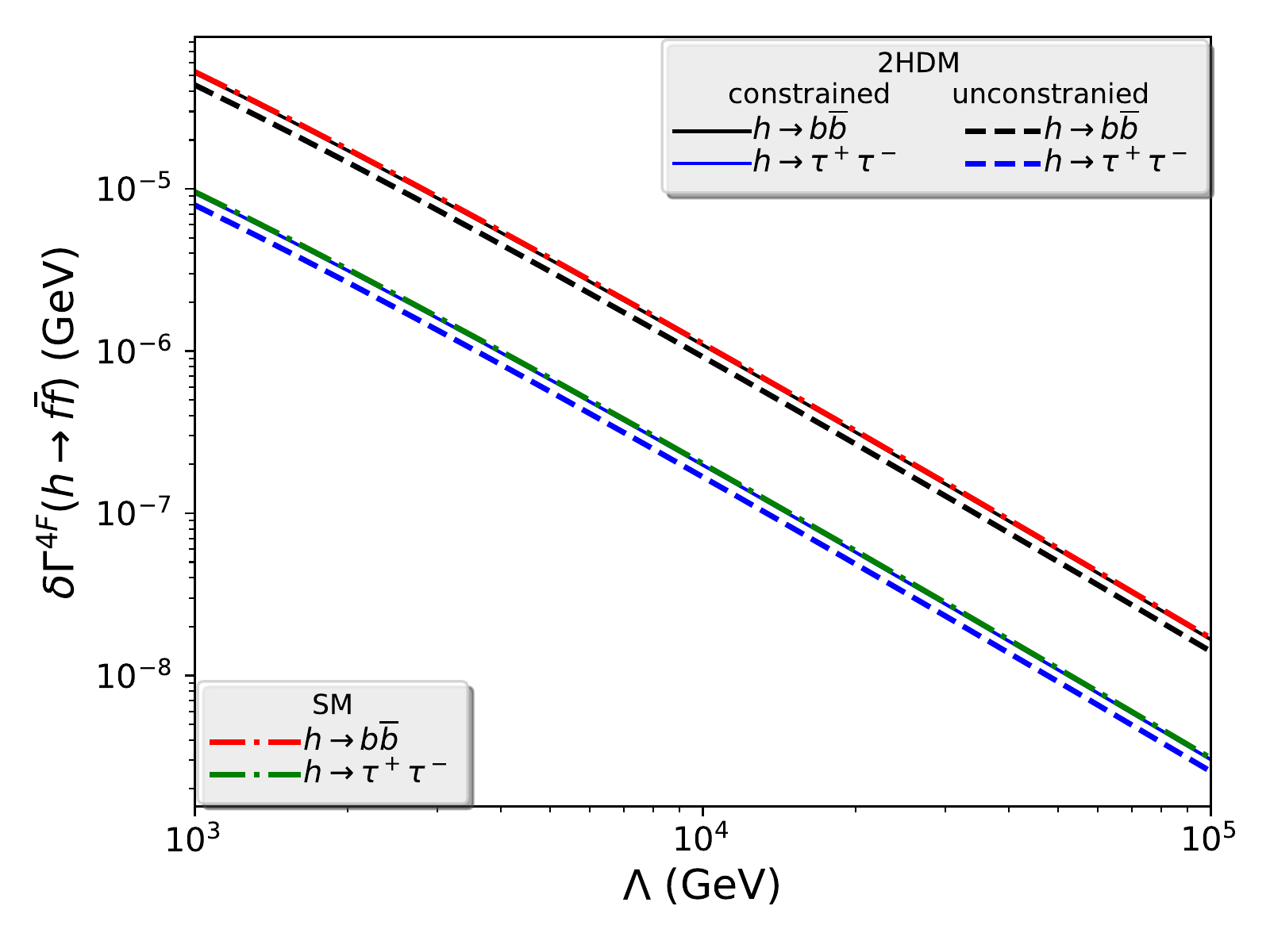}
  \end{center}
  \caption{Variation of the Higgs decay witdh into $b\bar{b}$ and $\tau\bar{\tau}$ at one-loop due to the 4-fermion interaction as a function of the new physics scale $\Lambda$.}
  \label{fig.widthHiggs2HDM}
\end{figure}

Considering the matching between the coupling constants in Eqs.~\eqref{eq.Lag} and \eqref{Yuk2HDM}, i.e. \(y_{s,p}^f~=~m_f \hat{y}_f^h/v\), we
put these values in our master formula for the scalar decays into both
\(b\bar{b}\) and \(\tau\bar{\tau}\). The variation of the Higgs partial
width decay due to the four-fermion effective interaction in the
context of the 2HDM are shown in Fig. \ref{fig.widthHiggs2HDM}. The
differences in the Higgs decay witdth variation between the SM and the
2HDM frameworks are small for all \(\Lambda\) (less than 1\%), because the deviations in
the yukawa coupling between the two cases are negligible.

\subsection{Decay width corrections to the heavy neutral (pseudo)scalars in the 2HDM}
\label{sec:orgf09b5a8}

Next, we estimate the corrections to the decay width to
the heavy neutral scalars \((H,A)\). We exemplify in the unconstrained Type II-1
model, whose Yukawa couplings are summarised in Table~\ref{tab.yuk_heavy}.

\begin{table}[htbp]
  \caption{\label{tab.yuk_heavy}
    Effective Yukawa couplings for Type II-1 unconstrained model for the massive scalar \(H^0\) and pseudoscalar \(A^0\) to fermions: up-type quarks, down-type quarks and charged leptons.}
  \centering
  \begin{tabular}{c|c|c}
    Yukawa coupling & Scalar (\(H^0\)) & Pseudoscalar (\(A^0\))\\
    \hline
    \(y_u\) & 2.69 & 2.77\\
    \(y_d\) & 0.37 & 0.36\\
    \(y_l\) & 0.37 & 0.36\\
  \end{tabular}
\end{table}

\begin{figure}[H]
  \includegraphics[width=.49\textwidth]{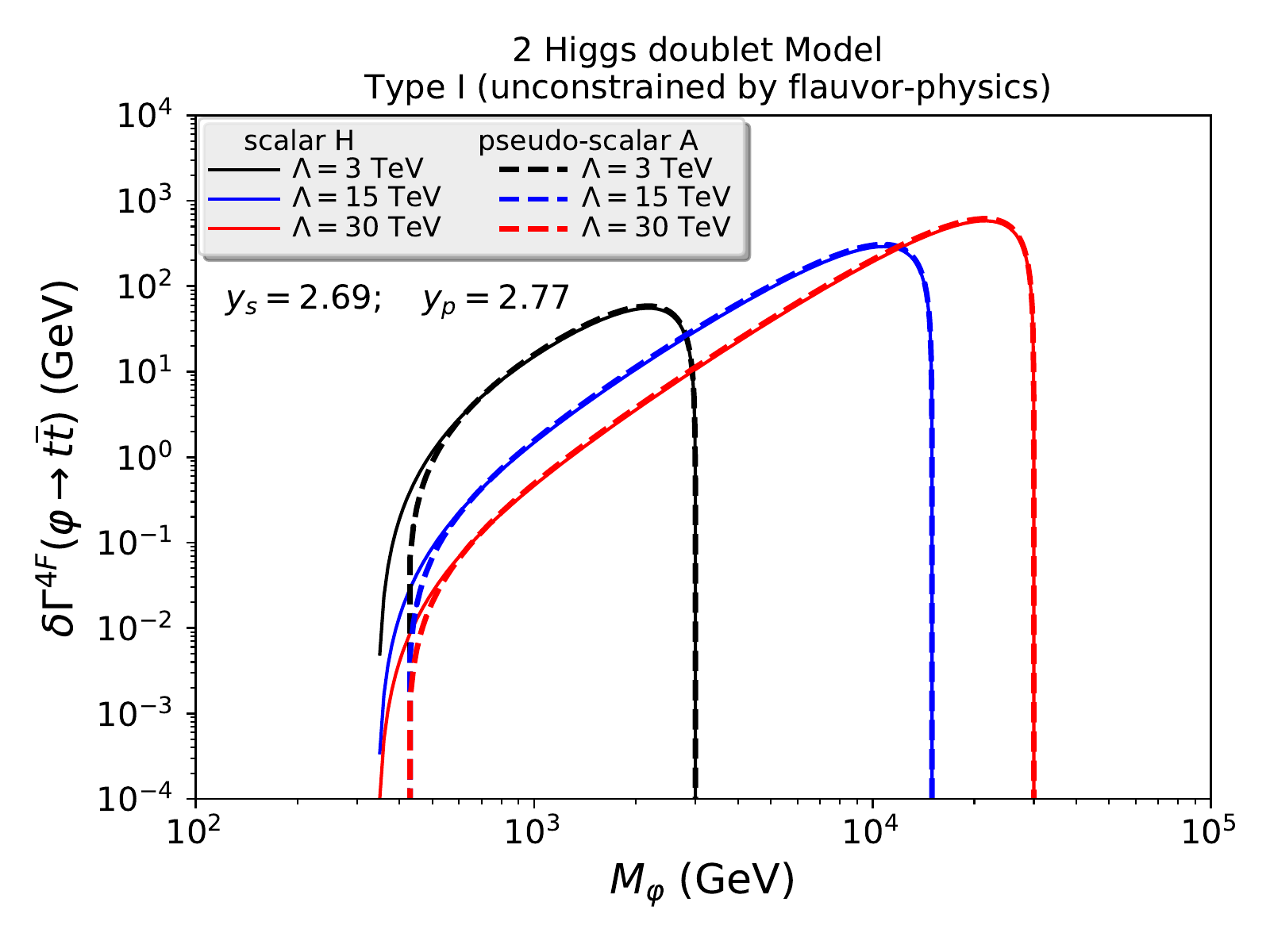}
  \includegraphics[width=.49\textwidth]{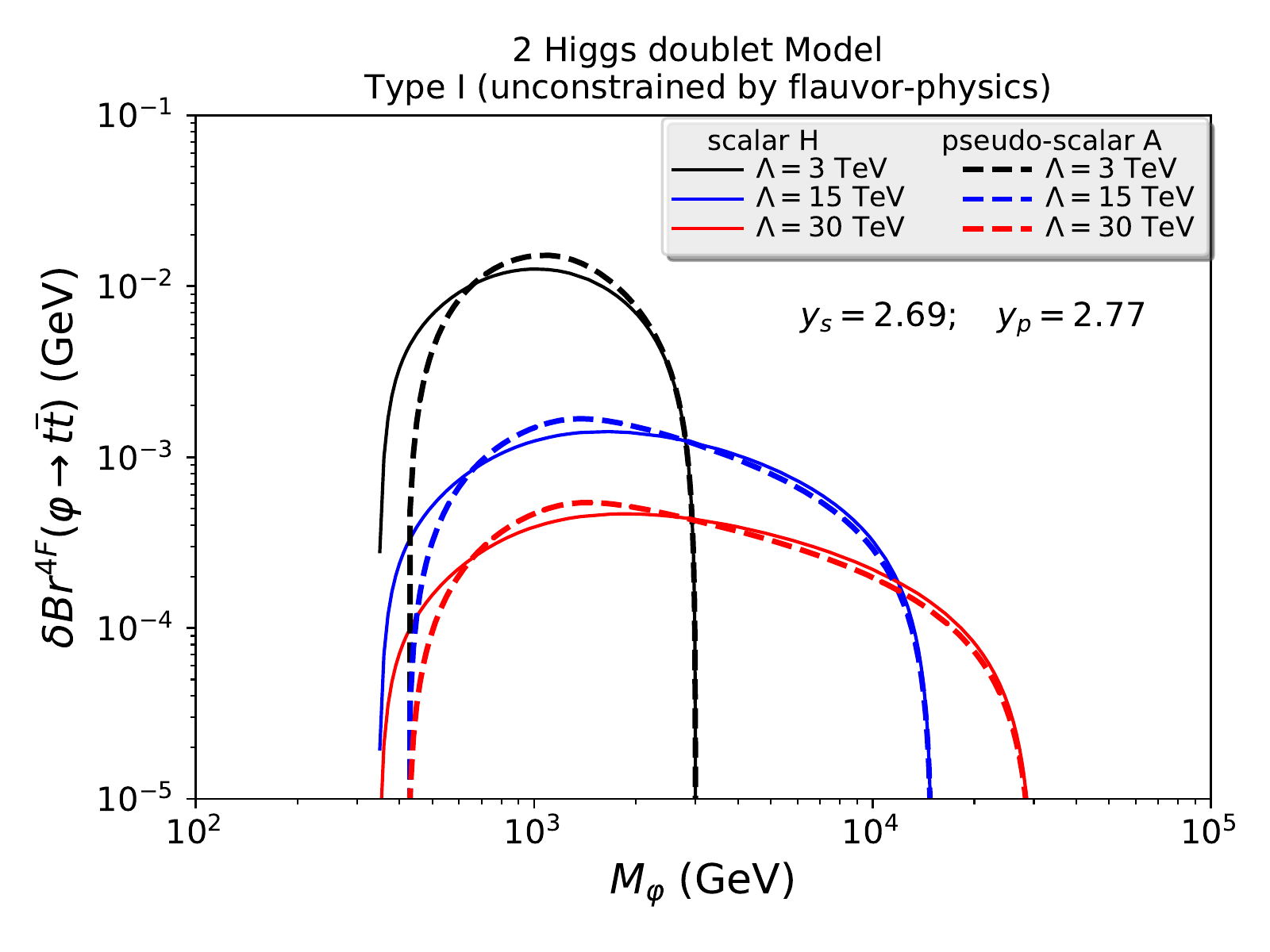}
  \newline\newline
  \hspace*{\stretch{1}} (a)\hspace{\stretch{2}}(b)\hspace{\stretch{1}}
  \caption{Variation of the decay widths (a) and branching ratios (b) into $t\bar{t}$ at one-loop for the heavy scalar $H$ and the pseudoscalar $A$ as a function of their mass.}
  \label{fig.widthHiggstoptop2HDM}
\end{figure}

The first important consequence to mention is that there is no important
difference in \(\delta\Gamma^{4F}\) between the scalar and pseudo-scalar case at any value of \(\Lambda\), except near the threshold. This is because at lower scalar masses there is a bigger suppresion in the pseudoscalar variation (see (18) and (19)), therefore making the scalar variation visible at lower scalar masses. Note that always the corrections are valid below $\Lambda$, which is the effective cut-off theory, and at \(M_\varphi = \Lambda\) the curves in the
plot fall steeply due to the logarithm behaviour in the correction. Complementary, Fig. \ref{fig.widthHiggstoptop2HDM}b shows the change in the branching fraction of the (pseudo)scalar to $t\bar{t}$ as a function of its mass, for the same cut-off values than before. As it is expected, the $\Lambda$ cuadratic suppresion in the (pseudo)scalar variation makes the corrections bigger for low gravitational scales ( $\sim 0.1$ to $1\%$), and suppressed for $\Lambda \gtrsim 15$ TeV, making a correction less than $0.1\%$. Note that for masses near the threshold, for any $\Lambda$, the branching fraction corrections in the scalar case can become as big as one order of magnitude than the pseudoscalar case one.

It is worth to point out that although our results for SM Higgs boson are less sensitive than the ones presented in Refs. \cite{Castillo-Felisola:2014xba} for Z boson decay, we cannot assure the same for the 2HDM model.

\section{Discussion and conclusions}
\label{sec.concl}
We have reviewed how gravitational torsion induces an effective
interaction between SM fermions. This new interactions affect directly
particle observables, such as their decay width. We analysed the
variation induced, by the torsion-descendent four-fermion interaction,
in scalar and pseudoscalar particles in the SM and the type II-1
2HDM.

Concerning to SM Higgs decays, we have focus on \(h \rightarrow
b\bar{b}\) and \(h \rightarrow \tau^+ \tau^-\) decays, which are the
dominant decay modes having branching ratios of
\SI{57}[\approx]{\percent} and \SI{6}[\approx]{\percent},
respectively. We have considered the correction to the branching ratio
for these processes mediated by the effective four-fermion interaction
at one-loop level. It can be seen in Fig. \ref{fig.Hdt}, that the
contribution to both fermionic channels become smaller as the
gravitational scale grows up. On the other hand,  \(\delta Br^{4F}(h
\rightarrow b\bar{b})\) is roughly speaking an order of magnitude
bigger than \(\delta Br^{4F}(h \rightarrow \tau^+\tau^-)\) for any scale energy
\(\Lambda\), doing this channel more relevant from a phenomenological
viewpoint. For gravitational scales as low as \(\Lambda =
\SI{1}{\TeV}\), the corrections to the branching ratio for \(h \rightarrow
b\bar{b}\) is \SI{1}[\sim]{\percent}, meanwhile \(h \rightarrow \tau^+
\tau^-\) is \SI{0.1}[\sim]{\percent}.  Moreover, from Fig. \ref{fig.Hdt}
(a), one can see that when \(\Lambda = \SI{30}{\TeV}\), the corrections
are \SI{3.3e-3}{\percent} for \(h \rightarrow b\bar{b}\) and
\SI{2.2e-4}{\percent} for \(h \rightarrow \tau^+ \tau^-\).

However, the \(b\bar{b}\) signal channel may be more visible at future
Higgs factories, such as the International Linear Collider (ILC) or
the Compact Linear Collider (CLIC), where the QCD background is
reduced, and therefore having more precision in some
observables. Additionally, it is expected higher precision measurements
in the Higgs sector at both ILC and CLIC than LHC , allowing to
explore more deeply into the quantitative information of the couplings
and Higgs decay width, and therefore being able to measure deviations
in the Higgs decay width, eventually as low as our results.

At this point, we want to remark one more time that our results have shown that the Higgs decay width is less sensitive than, for instance, the \(Z\) boson decay width
\cite{Castillo-Felisola:2014xba} to the kind of corrections we are studying. This is
mainly due to the higher number of degrees of freedom present in the
vector case and to the fact that the properties of the \(Z\) boson have
been measured with a high accuracy.

On the other hand, our results turn out to be more auspicious in the
case of the 2HDM, particularly if the non-standard scalars are heavy,
as shown if Fig. \ref{fig.widthHiggstoptop2HDM}. It is even possible to
distinguish between scalars and pseudo-scalars near the threshold of
the decay channel if there are additional heavy fermions. The
corrections \(\delta\Gamma_{\text{4FI}}^{\text{P}}\) and \(\delta
\Gamma_{\text{ 4FI}}^{\text{ S}}\) can be distinguished in the lower
mass threshold, when we have provided \(y_s = 2.69\) and \(y_p =
2.77\). However, it is important to note that in general (arbitrary
values of \(y_s\) and \(y_p\)) the condition of distinguishability is
\begin{equation}
  y_{p}^2 \neq y_{s}^2 \, \frac{M_{\phi}^2 - 2
    m_{f}^2}{M_{\phi}^2 - 6 m_{f}^2} \; \Bigg( 1 - \frac{4
    m_{f}^2}{M_{\phi}^2} \Bigg).
\end{equation}

\section*{Acknowledgments}

B.D.S. thanks to the Universidad Austral de Chile for the hospitality
during the completion of this work, to Conicyt Becas-Chile and PIIC/DGIP for the support. F.R. thanks to the University of
Southampton (UK) for the hospitality during the completion of this
work. This work was partially supported by Fellowship Grant \emph{Becas
  Chile} No. 74160012, USM grant No. 11.15.77, CONICYT (Chile) under
project No. 79140040 and FONDECYT (Chile)  grant No.  1160423, PIA/Basal FB0821 and Conicyt ACT1406. The Centro Cient\'ifico Tecnol\'ogico de Valpara\'iso
(CCTVal) is funded by the Chilean Government through the Centers of
Excellence Basal Financing Program FB0821 of CONICYT.

\bibliographystyle{unsrt}
\bibliography{References}

\end{document}

%% file: Def-org.tex
\usepackage{mathtools,amsmath,amssymb,amsfonts,dsfont,mathrsfs,amsthm,latexsym}

\usepackage{bm}
\usepackage{graphicx}
\usepackage{centernot}
\usepackage{xcolor}
\usepackage{comment}
\usepackage{feynmf}
\usepackage{siunitx}
\usepackage{array}
\newcolumntype{L}[1]{>{\raggedright\let\newline\\\arraybackslash\hspace{0pt}}m{#1}}
\newcolumntype{C}[1]{>{\centering\let\newline\\\arraybackslash\hspace{0pt}}m{#1}}
\newcolumntype{R}[1]{>{\raggedleft\let\newline\\\arraybackslash\hspace{0pt}}m{#1}}
\usepackage{tikz}
\usepackage{braket}
\usepackage{xparse}
\usepackage{breqn}
\usetikzlibrary{shapes,
  snakes,
  decorations.pathmorphing,
  decorations.markings,
  calc,
  shadows.blur,
  shadings}
\usepackage[framemethod=tikz]{mdframed}

\makeatletter
\pgfdeclaredecoration{gluon}{coil}
{
  \state{coil}[switch if less than=%
    0.5\pgfdecorationsegmentlength+
    \pgfdecorationsegmentaspect\pgfdecorationsegmentamplitude+%
    \pgfdecorationsegmentaspect\pgfdecorationsegmentamplitude to last,
               width=+\pgfdecorationsegmentlength]
  {
    \pgfpathcurveto
    {\pgfpoint@oncoil{0    }{ 0.555}{1}}
    {\pgfpoint@oncoil{0.445}{ 1    }{2}}
    {\pgfpoint@oncoil{1    }{ 1    }{3}}
    \pgfpathcurveto
    {\pgfpoint@oncoil{1.555}{ 1    }{4}}
    {\pgfpoint@oncoil{2    }{ 0.555}{5}}
    {\pgfpoint@oncoil{2    }{ 0    }{6}}
    \pgfpathcurveto
    {\pgfpoint@oncoil{2    }{-0.555}{7}}
    {\pgfpoint@oncoil{1.555}{-1    }{8}}
    {\pgfpoint@oncoil{1    }{-1    }{9}}
    \pgfpathcurveto
    {\pgfpoint@oncoil{0.445}{-1    }{10}}
    {\pgfpoint@oncoil{0    }{-0.555}{11}}
    {\pgfpoint@oncoil{0    }{ 0    }{12}}
  }
  \state{last}[next state=final]
  {
    \pgfpathcurveto
    {\pgfpoint@oncoil{0    }{ 0.555}{1}}
    {\pgfpoint@oncoil{0.445}{ 1    }{2}}
    {\pgfpoint@oncoil{1    }{ 1    }{3}}
    \pgfpathcurveto
    {\pgfpoint@oncoil{1.555}{ 1    }{4}}
    {\pgfpoint@oncoil{2    }{ 0.555}{5}}
    {\pgfpoint@oncoil{2    }{ 0    }{6}}
  }
  \state{final}{}
}

\def\pgfpoint@oncoil#1#2#3{%
  \pgf@x=#1\pgfdecorationsegmentamplitude%
  \pgf@x=\pgfdecorationsegmentaspect\pgf@x%
  \pgf@y=#2\pgfdecorationsegmentamplitude%
  \pgf@xa=0.083333333333\pgfdecorationsegmentlength%
  \advance\pgf@x by#3\pgf@xa%
}
\makeatother

\tikzset{
  boson/.style={decorate,decoration={gluon,segment length=9pt,aspect=0}},
  on each segment/.style={
    decorate,
    decoration={
      show path construction,
      moveto code={},
      lineto code={
        \path [#1]
        (\tikzinputsegmentfirst) -- (\tikzinputsegmentlast);
      },
      curveto code={
        \path [#1] (\tikzinputsegmentfirst)
        .. controls
        (\tikzinputsegmentsupporta) and (\tikzinputsegmentsupportb)
        ..
        (\tikzinputsegmentlast);
      },
      closepath code={
        \path [#1]
        (\tikzinputsegmentfirst) -- (\tikzinputsegmentlast);
      },
    },
  },
  mid arrow/.style={postaction={decorate,decoration={
        markings,
        mark=at position .5 with {\arrow[#1]{stealth}}
      }}},
}
\tikzset{
    vector/.style={decorate, decoration={snake}, draw},
	provector/.style={decorate, decoration={snake,amplitude=2.5pt}, draw},
	antivector/.style={decorate, decoration={snake,amplitude=-2.5pt}, draw},
    fermion/.style={draw=black, postaction={decorate},
        decoration={markings,mark=at position .55 with {\arrow[draw=black]{>}}}},
    fermionbar/.style={draw=black, postaction={decorate},
        decoration={markings,mark=at position .55 with {\arrow[draw=black]{<}}}},
    fermionnoarrow/.style={draw=black},
    gluon/.style={decorate, draw=black,
        decoration={coil,amplitude=4pt, segment length=5pt}},
    scalar/.style={dashed,draw=black, postaction={decorate},
        decoration={markings,mark=at position .55 with {\arrow[draw=black]{>}}}},
    scalarbar/.style={dashed,draw=black, postaction={decorate},
        decoration={markings,mark=at position .55 with {\arrow[draw=black]{<}}}},
    scalarnoarrow/.style={dashed,draw=black},
    electron/.style={draw=black, postaction={decorate},
        decoration={markings,mark=at position .55 with {\arrow[draw=black]{>}}}},
	bigvector/.style={decorate, decoration={snake,amplitude=4pt}, draw},
}

\DeclareDocumentCommand{\Ag}{ s o }{ \IfBooleanTF{#1}
    { \IfValueTF{#2}{ \bm{\mathcal{A}}_{(#2)} }{ \bm{\mathcal{A}} } }
    { \IfValueTF{#2}{    {\mathcal{A}}_{(#2)} }{    {\mathcal{A}} } } }
\DeclareDocumentCommand{\Af}{ s o }{ \IfBooleanTF{#1}
    { \IfValueTF{#2}{ \boldsymbol{A}_{(#2)} }{ \boldsymbol{A} } }
    { \IfValueTF{#2}{            {A}_{(#2)} }{            {A} } } }

\newcommand{\cdf}[1][]{{\boldsymbol{\mathcal{D}}}{#1}}

\newcommand{\df}[1][]{\mathbf{d}{#1}}

\DeclareDocumentCommand{\Fg}{ s o }{ \IfBooleanTF{#1}
    { \IfValueTF{#2}{ \bm{\mathcal{F}}_{(#2)} }{ \bm{\mathcal{F}} } }
    { \IfValueTF{#2}{    {\mathcal{F}}_{(#2)} }{    {\mathcal{F}} } } }
\DeclareDocumentCommand{\Ff}{ s o }{ \IfBooleanTF{#1}
    { \IfValueTF{#2}{ \boldsymbol{F}_{(#2)} }{ \boldsymbol{F} } }
    { \IfValueTF{#2}{            {F}_{(#2)} }{            {F} } } }

\newcommand{\ga}{\gamma}
\newcommand{\gf}{\boldsymbol{\gamma}}

\DeclareMathOperator{\st}{\star}

\newcommand{\Lag}{\mathscr{L}}

\DeclareDocumentCommand{\PB}{ O{m} O{q} O{p} m m }{ \frac{ \partial #4 }{\partial {#2}^{#1} } \frac{ \partial #5 }{\partial {#3}_{#1} } - \frac{ \partial #4 }{\partial {#3}_{#1} } \frac{ \partial #5 }{\partial {#2}^{#1} } }

\DeclareDocumentCommand\Te{o o m }{\mathcal{T}{}^{#1}_{#2}(#3)}

\newcommand{\we}{{\scriptstyle\wedge}}

\DeclareDocumentCommand{\BM}{ s }{ \IfBooleanTF{#1} {\hat{\bm{M}}}{\bm{M}} }
\DeclareDocumentCommand{\BN}{ s }{ \IfBooleanTF{#1} {\hat{\bm{N}}}{\bm{N}} }
\DeclareDocumentCommand{\BP}{ s }{ \IfBooleanTF{#1} {\hat{\bm{P}}}{\bm{P}} }
\DeclareDocumentCommand{\BQ}{ s }{ \IfBooleanTF{#1} {\hat{\bm{Q}}}{\bm{Q}} }
\DeclareDocumentCommand{\BR}{ s }{ \IfBooleanTF{#1} {\hat{\bm{R}}}{\bm{R}} }
\DeclareDocumentCommand{\BS}{ s }{ \IfBooleanTF{#1} {\hat{\bm{S}}}{\bm{S}} }
\DeclareDocumentCommand{\BU}{ s }{ \IfBooleanTF{#1} {\hat{\bm{U}}}{\bm{U}} }
\DeclareDocumentCommand{\BV}{ s }{ \IfBooleanTF{#1} {\hat{\bm{V}}}{\bm{V}} }

\NewDocumentCommand\MyAc{ m }{#1}
\DeclareDocumentCommand{\vif}{ t. t, t- s s m }{
  \RenewDocumentCommand\MyAc{ m }{##1}
  \IfBooleanT{#1}{\RenewDocumentCommand\MyAc{ m }{ \mathring{##1} } }
  \IfBooleanT{#2}{\RenewDocumentCommand\MyAc{ m }{ \tilde{##1} } }
  \IfBooleanT{#3}{\RenewDocumentCommand\MyAc{ m }{ \bar{##1} } }
  \IfBooleanTF{#4}
  { \IfBooleanTF{#5} { \hat{\MyAc{\boldsymbol{e}}}^{\hat{#6}} }{ \hat{\MyAc{\boldsymbol{e}}}^{{#6}} } }
  { \MyAc{\boldsymbol{e}}^{{#6}} } }
\DeclareDocumentCommand{\vi}{ t. t, t- s s m m}{
  \RenewDocumentCommand\MyAc{ m }{##1}
  \IfBooleanT{#1}{\RenewDocumentCommand\MyAc{ m }{ \mathring{##1} } }
  \IfBooleanT{#2}{\RenewDocumentCommand\MyAc{ m }{ \tilde{##1} } }
  \IfBooleanT{#3}{\RenewDocumentCommand\MyAc{ m }{ \bar{##1} } }
  \IfBooleanTF{#4}
  { \IfBooleanTF{#5} { \hat{\MyAc{e}}^{\hat{#6}}_{\hat{#7}} }{ \hat{\MyAc{e}}^{#6}_{{#7}} } }
  { \MyAc{e}^{{#6}}_{{#7}} } }

\DeclareDocumentCommand{\bt}{ t. t, t- s s m m m }{
  \RenewDocumentCommand\MyAc{ m }{##1}
  \IfBooleanT{#1}{\RenewDocumentCommand\MyAc{ m }{ \mathring{##1} } }
  \IfBooleanT{#2}{\RenewDocumentCommand\MyAc{ m }{ \tilde{##1} } }
  \IfBooleanT{#3}{\RenewDocumentCommand\MyAc{ m }{ \bar{##1} } }
  \IfBooleanTF{#4}
  { \IfBooleanTF{#5} { \hat{\MyAc{B}}_{{#6}}{}^{\hat{#7}}{}_{\hat{#8}} }{ \hat{\MyAc{\Gamma}}_{{#6}}{}^{{#7}}{}_{{#8}} } }
  { \MyAc{B}_{{#6}}{}^{{#7}}{}_{{#8}} } }

\DeclareDocumentCommand{\ct}{ t. t, t- s s m m m }{
  \RenewDocumentCommand\MyAc{ m }{##1}
  \IfBooleanT{#1}{\RenewDocumentCommand\MyAc{ m }{ \mathring{##1} } }
  \IfBooleanT{#2}{\RenewDocumentCommand\MyAc{ m }{ \tilde{##1} } }
  \IfBooleanT{#3}{\RenewDocumentCommand\MyAc{ m }{ \bar{##1} } }
  \IfBooleanTF{#4}
  { \IfBooleanTF{#5} { \hat{\MyAc{\Gamma}}_{{#6}}{}^{\hat{#7}}{}_{\hat{#8}} }{ \hat{\MyAc{\Gamma}}_{{#6}}{}^{{#7}}{}_{{#8}} } }
  { \MyAc{\Gamma}_{{#6}}{}^{{#7}}{}_{{#8}} } }
\DeclareDocumentCommand{\spif}{ t. t, t- s s m m }{
  \RenewDocumentCommand\MyAc{ m }{##1}
  \IfBooleanT{#1}{\RenewDocumentCommand\MyAc{ m }{ \mathring{##1} } }
  \IfBooleanT{#2}{\RenewDocumentCommand\MyAc{ m }{ \tilde{##1} } }
  \IfBooleanT{#3}{\RenewDocumentCommand\MyAc{ m }{ \bar{##1} } }
  \IfBooleanTF{#4}
  { \IfBooleanTF{#5} { \hat{\MyAc{\boldsymbol{\omega}}}^{\hat{#6}}{}_{\hat{#7}} }{ \hat{\MyAc{\boldsymbol{\omega}}}^{{#6}}{}_{{#7}} } }
  { \MyAc{\boldsymbol{\omega}}^{{#6}}{}_{{#7}} } }
\DeclareDocumentCommand{\spi}{ t. t, t- s s m m m }{
  \RenewDocumentCommand\MyAc{ m }{##1}
  \IfBooleanT{#1}{\RenewDocumentCommand\MyAc{ m }{ \mathring{##1} } }
  \IfBooleanT{#2}{\RenewDocumentCommand\MyAc{ m }{ \tilde{##1} } }
  \IfBooleanT{#3}{\RenewDocumentCommand\MyAc{ m }{ \bar{##1} } }
  \IfBooleanTF{#4}
  { \IfBooleanTF{#5} { \hat{\MyAc{{\omega}}}_{\hat{#6}}{}^{\hat{#7}}{}_{\hat{#8}} }{ \hat{\MyAc{{\omega}}}_{{#6}}{}^{{#7}}{}_{{#8}} } }
  { \MyAc{{\omega}}_{{#6}}{}^{{#7}}{}_{{#8}} } }

\DeclareDocumentCommand{\rif}{ t. t, t- s s m m }{
  \RenewDocumentCommand\MyAc{ m }{##1}
  \IfBooleanT{#1}{\RenewDocumentCommand\MyAc{ m }{ \mathring{##1} } }
  \IfBooleanT{#2}{\RenewDocumentCommand\MyAc{ m }{ \tilde{##1} } }
  \IfBooleanT{#3}{\RenewDocumentCommand\MyAc{ m }{ \bar{##1} } }
  \IfBooleanTF{#4}
  { \IfBooleanTF{#5} { \hat{\MyAc{\bm{\mathcal{R}}}}{}^{\hat{#6}}{}_{\hat{#7}} }{ \hat{\MyAc{\bm{\mathcal{R}}}}{}^{{#6}}{}_{{#7}} } }
  { \MyAc{\bm{\mathcal{R}}}{}^{{#6}}{}_{{#7}} } }
\DeclareDocumentCommand{\ri}{ t. t, t- s s m m m }{
  \RenewDocumentCommand\MyAc{ m }{##1}
  \IfBooleanT{#1}{\RenewDocumentCommand\MyAc{ m }{ \mathring{##1} } }
  \IfBooleanT{#2}{\RenewDocumentCommand\MyAc{ m }{ \tilde{##1} } }
  \IfBooleanT{#3}{\RenewDocumentCommand\MyAc{ m }{ \bar{##1} } }
  \IfBooleanTF{#4}
  { \IfBooleanTF{#5} { \hat{\MyAc{\mathcal{R}}}_{{#6}}{}^{\hat{#7}}{}_{\hat{#8}} }{ \hat{\MyAc{\mathcal{R}}}_{{#6}}{}^{{#7}}{}_{{#8}} } }
  { \MyAc{\mathcal{R}}_{{#6}}{}^{{#7}}{}_{{#8}} } }

\newcommand{\Rif}[2]{\bm{\mathcal{R}}^{{#1}}{}_{{#2}}}

\DeclareDocumentCommand{\kf}{ t. t, t- s s m m }{
  \RenewDocumentCommand\MyAc{ m }{##1}
  \IfBooleanT{#1}{\RenewDocumentCommand\MyAc{ m }{ \mathring{##1} } }
  \IfBooleanT{#2}{\RenewDocumentCommand\MyAc{ m }{ \tilde{##1} } }
  \IfBooleanT{#3}{\RenewDocumentCommand\MyAc{ m }{ \bar{##1} } }
  \IfBooleanTF{#4}
  { \IfBooleanTF{#5} { \hat{\MyAc{\bm{\mathcal{K}}}}^{\hat{#6}}{}_{\hat{#7}} }{ \hat{\MyAc{\bm{\mathcal{K}}}}^{{#6}}{}_{{#7}} } }
  { \MyAc{\bm{\mathcal{K}}}^{{#6}}{}_{{#7}} } }
\DeclareDocumentCommand{\ko}{ t. t, t- s s m m m }{
  \RenewDocumentCommand\MyAc{ m }{##1}
  \IfBooleanT{#1}{\RenewDocumentCommand\MyAc{ m }{ \mathring{##1} } }
  \IfBooleanT{#2}{\RenewDocumentCommand\MyAc{ m }{ \tilde{##1} } }
  \IfBooleanT{#3}{\RenewDocumentCommand\MyAc{ m }{ \bar{##1} } }
  \IfBooleanTF{#4}
  { \IfBooleanTF{#5} { \hat{\MyAc{\mathcal{K}}}_{\hat{#6}}{}^{\hat{#7}}{}_{\hat{#8}} }{ \hat{\MyAc{\mathcal{K}}}_{{#6}}{}^{{#7}}{}_{{#8}} } }
  { \MyAc{\mathcal{K}}_{{#6}}{}^{{#7}}{}_{{#8}} } }

\newcommand{\hcont}[3]{\hat{\mathcal{K}}_{#1}{}^{#2}{}_{#3}}

\DeclareDocumentCommand{\tf}{ t. t, t- s s m }{
  \RenewDocumentCommand\MyAc{ m }{##1}
  \IfBooleanT{#1}{\RenewDocumentCommand\MyAc{ m }{ \mathring{##1} } }
  \IfBooleanT{#2}{\RenewDocumentCommand\MyAc{ m }{ \tilde{##1} } }
  \IfBooleanT{#3}{\RenewDocumentCommand\MyAc{ m }{ \bar{##1} } }
  \IfBooleanTF{#4}
  { \IfBooleanTF{#5} { \hat{\MyAc{\bm{\mathcal{T}}}}^{\hat{#6}} }{ \hat{\MyAc{\bm{\mathcal{T}}}}^{{#6}} } }
  { \MyAc{\bm{\mathcal{T}}}^{{#6}} } }
\newcommand{\tor}{\mathcal{T}}
\newcommand{\tors}[3]{\mathcal{T}{}_{#1}{}^{#2}{}_{#3}}

\newcommand{\Tf}[1]{\bm{\mathcal{T}}^{#1}}

\newcommand{\beq}{\begin{equation}}
\newcommand{\eeq}{\end{equation}}
\newcommand{\ber}{\begin{eqnarray}}
\newcommand{\eer}{\end{eqnarray}}

\NewDocumentCommand{\tak}{ s m m}{
  \IfBooleanTF{#1}{ \big( {#2} \big) \big[ {#3} \big] }
              { \big( {#2} \big] \big[ {#3} \big) }
}

\newcommand*{\de}[1]{\mathop{\mathrm{d}#1}\nolimits}

\mdfdefinestyle{ibox}{
  align = center,
  linecolor=green!60!black, 
  outerlinewidth = 1pt,
  frametitle={Infobox},
  frametitlerule=true,
  frametitlebackgroundcolor=green!15,
  backgroundcolor=green!10,
  roundcorner=8pt,
  frametitlealignment=\centering,
}
\newmdenv[style=ibox]{infobox}

\mdfdefinestyle{theoremstyle}{%
  linecolor=red!40,
  linewidth=2pt,%
  frametitlerule=true,%
  frametitlebackgroundcolor=gray!20,
  innertopmargin=\topskip,
}
\mdtheorem[style=theoremstyle]{cdbexample}{Cadabra Example}

\newcommand\UTFSM{Departamento de F\'isica, Universidad T\'{e}cnica Federico Santa Mar\'\i a, Casilla 110-V, Valpara\'iso, Chile}

\newcommand\CCTVal{Centro Cient\'ifico Tecnol\'ogico de Valpara\'iso, Casilla 110-V, Valpara\'\i so, Chile}

\newcommand\DLNP{Dzhelepov Laboratory of Nuclear Problems, Joint Institute for Nuclear Research,\\ 141980 Dubna, Moscow Region, Russian Federation}